\documentclass[pre,twocolumn,aps,showpacs]{revtex4}
\usepackage{graphicx}
\usepackage[activeacute,english]{babel}
\begin{document}

\title {   
Heat fluctuations in Brownian transducers 
}
\author{
A. Gomez-Marin
}
\author{
J. M. Sancho
}
\affiliation{
Departament d'Estructura i Constituents de la Materia,
Facultat de Fisica, Universitat de Barcelona, Diagonal 647, 08028 Barcelona,
Spain} 

\date{\today}
     
\begin{abstract} 
Heat fluctuation probability distribution function in Brownian
transducers operating between two heat reservoirs is studied.
We find, both analytically and numerically, that 
the recently proposed Fluctuation Theorem for Heat Exchange [C. Jarzynski
and D. K. Wojcik, Phys. Rev. Lett. {\bf 92}, 230602 (2004)] has to be modified 
when the coupling mechanism between both baths is considered.
We also extend such relation when external work is present.
\end{abstract}


\maketitle

Nonequilibrium systems are receiving much 
attention from a theoretical point of view through the derivation of 
the so called fluctuation theorems.
The theoretical approach is based on microscopic reversibility and  
elegant analytical properties for the
probability distribution of entropy generated are derived
\cite{croo}.
From these rigorous results, 
corollaries such as a statistical derivation of the Second Law can be
achieved. 
Moreover, they hold for systems arbitrarily far
away from equilibrium and are not restricted to the linear regime.
There are different fluctuation relations depending on the 
dynamics they apply to, the magnitudes they relate or the state of the system
they refer to. Amongst the most relevant one finds 
the Gallavoti-Cohen fluctuation theorem \cite{gc},
the Jarzynski equality \cite{df}, 
the formalism for steady-state thermodynamics \cite{sst},
an extension of the fluctuation theorem \cite{c1}
and an integral fluctuation theorem \cite{est}. Apart from their intrinsic
value to theoretical physics, a vast
number of experimental applications based on such results have been
developed \cite{rit}. The benchmark of these 
theorems are nanosystems such as molecular motors \cite{phto}.
Nearly all of such relations focus on work fluctuations
\cite{maz,c3,dw}. In contrast, not much attention has been paid to heat 
fluctuations. There are only a few contributions to the topic
\cite{c1,c2,hh}, in which only one thermal bath is considered.
Recently, Jarzynski and Wojcik \cite{jar} derived a fluctuation
theorem  for heat exchange (denoted as XFT) between to systems initially
prepared at 
different temperatures $T_{1}$ and $T_{2}$, then  placed in thermal contact
with one another for a certain time and, finally, separated again. The theorem 
states that the probability distribution $p(Q)$ of the heat exchange 
$Q$ satisfies
\begin{equation} \label{xft!}
\ln \frac{p(+Q)}{p(-Q)}=\Delta \beta Q,
\end{equation}
where $\Delta \beta=1/T_{1}-1/T_{2}$, and  $k_{B}=1$ is taken 
for simplicity from now on. Both systems must be prepared in equilibrium and
then placed in thermal contact with one another, for an arbitrary period of
time $t_0$, during which a net heat $Q$ is transferred.
This is a very interesting result because, in the first
place, it focus on heat fluctuations and, secondly, because of its universal
character: the prediction depends only on the two temperatures and not on
any characteristic of the systems.  
Nevertheless, the theorem was derived for the presence of heat exchange only,
without considering other sources of energy such as external work or
the energy involved in the connecting mechanisms.   
Then, it is supposed that the heat lost by one system is exactly compensated
the amount energy gained by the other.
However, any two bodies put in contact need a mechanism that connects them
and therefore an interaction term should be considered and its relevance and
effects studied. 
We may also notice that when work comes into play, the
transfer of heat must be revised because, according the Second Law,
both baths interchange different amounts of heat.

The purpose of this work is to study the predictions and applicability of
the XFT in specific models, such as Brownian motors. For a simple mechanical
system that allows heat exchange between two baths, we show that the XFT
has to be modified and we present an analytical calculation of such
modification, which does not depend on the details of the connecting
mechanism. 
We also propose an extension for 
the case in which external work is present.

\begin{figure}
\begin{center}
  \includegraphics[ angle=0, width=0.42\textwidth]{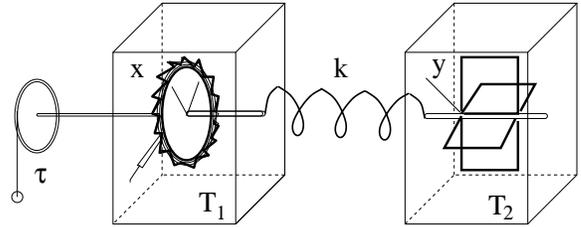}
  \caption{Ratchet, pawl and spring Brownian motor.}
  \label{rps}
\end{center}
\end{figure}
Brownian motors are a set of particularly peculiar
machines that make use of thermal fluctuations of the environment they are
immersed in to perform useful work \cite{reim}. There are many different
models being
the Feynman ratchet and pawl device \cite{feyn} the paradigm.
In general, Feynman--like Brownian motors are built from two subsystems
immersed respectively in  two thermal baths at different
temperatures and connected by some mechanism.
These engines, such as the motor in Ref. \cite{sbm},
fit perfectly the study of heat fluctuations from the point of view
of the setting of the XFT. Let's consider the transducer in
Fig. \ref{rps}. It has two degrees of freedom, 
$x$ and $y$, at different temperatures $T_{1} < T_{2}$ and connected through
a spring. Note that in the colder bath there is 
a saw-toothed wheel which acts as a ratchet potential.
The Langevin equations of motion of this device in the overdamped limit, when 
setting the friction equal to one, are
\begin{equation} \label{123}
\dot{x}=-\partial_{x} V_c(x,y) - V_{r}'(x) + \tau +\xi_{1}(t),
\end{equation}
\begin{equation} \label{124}
\dot{y}=-\partial_{y} V_c(x,y)+\xi_{2}(t),
\end{equation}
where $V_c(x,y)=(k/2)(x-y)^{2}$ is a harmonic coupling, $V_r(x)$ is the
ratchet potential in \cite{sbm} and $\tau$ is an external load.
$\xi_{1}(t)$ and $\xi_{2}(t)$ are independent  Gaussian white noises of
correlation  $\langle \xi_i(t) \xi_j(t') \rangle = 2 T_i \delta_{ij}
\delta(t-t') $.

\begin{figure}
\begin{center}
  \includegraphics[ angle=270, width=0.38\textwidth]{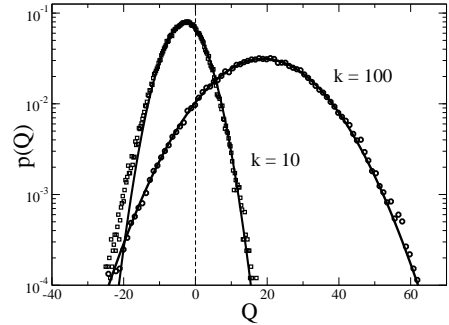}
  \caption{Histograms of the heat transfer $Q$ for the model
    (\ref{123})--(\ref{124}) at nonzero
   external torque for a Gaussian and a non-Gaussian case. The parameters are
   $\tau=5$, $V_{0}=1$, $d=12$, $t_{0}=1$,
   $T_{1}=1$ and $T_{2}=1.5$.
   The continuous lines are Gaussian fits.} 
  \label{one}
\end{center}
\end{figure}

Heat will flow from one bath to the other through the spring.
We call $Q$ the heat outcoming from the reservoir at $T_2$.
Note that due to the internal energy stored in
the spring, in the ratchet potential and due to the mechanical work,
$Q$ is not equal to the heat dissipated in the thermal 
bath at $T_1$. 
From Sekimoto's energetics scheme \cite{seki} one can obtain $Q$ as
\begin{equation} \label{dotQ}
Q= \int_{y(0)}^{y(t_{0})} ( \xi_{2}(t) - \dot{y}(t) ) dy(t) = k
\int_{0}^{t_{0}} (y(t)-x(t)) \dot{y}(t)  dt.  
\end{equation}
Before a formal approach to the analytic properties of this central quantity,
let's perform a numerical study of it.  
A numerical computation of the stochastic heat $Q$ is easily done by 
simulating the dynamics of the motor and performing the above integral
numerically, once the steady state has been reached. Then we can get a 
collection of values of the total heat $Q$ transferred during a fixed time
interval $t_{0}$, with which histograms can be built.
This is not exactly the situation described in the derivation of the XFT. 
According to Ref. \cite{jar}, both subsystems are initially in
equilibrium, then connected for a time period $t_{0}$ and, finally, 
separated again. However, in our simulations and calculations the transient
regime hardly contributes compared to the steady state state. 
Preliminary simulations in the nonlinear Brownian transducer at nonzero
external work conditions show that, in the limit $kt_{0}>>1$,
heat histograms are very well fitted by Gaussian distributions.
For smaller $kt_{0}$ distributions deviate from Gaussianity (see
Fig. \ref{one}). Several numerical explorations performed with different values
of the parameters involved, measuring the kurtosis and the skewness of the
histograms and checking the tails of the distributions reveal a wide range of
the parameter space in which the Gaussian approximation is justified. 
Therefore, in an appropriate and also physical limit, we can write
\begin{equation}  \label{gaussian}
 p(Q) = \frac{1}{(2\pi \sigma^2)^{1/2}} \; e^{-\frac{(Q-\langle Q
    \rangle)^2}{2 \sigma^2}} 
\end{equation}
and, accordingly, the XFT (\ref{xft!}) can be expressed as,
\begin{equation}
\ln \frac{p(+Q)}{p(-Q)} = Q \frac{2 \langle Q \rangle}{ \sigma^2}. 
\label{xftgauss}
\end{equation}

This is a very interesting situation because, if we were able to calculate 
the quantities $\langle Q \rangle$ and $ \sigma^2 = \langle \Delta Q ^2
\rangle$  for any specific model, then we could test the XFT
prediction on a non-ideal case. This cannot be done analytically for the
nonlinear 
model (\ref{123})--(\ref{124}) but it is possible if we simplify it by taking
$V_r=0$. In particular, for this 
passive Brownian transducer and in the absence of external load ($\tau=0$) we
find 
\begin{equation} \label{2}
\langle Q \rangle =t_{0}\frac{k}{2}(T_{2}-T_{1}),
\end{equation}
which is indeed Fourier's law for the thermal conductivity (see
Refs. \cite{parr,broeck}) and,
in the limit $kt_{0}>>1$, the second moment is \footnote{The calculation of
  the moments of $Q$ is tedious. We rewrite equation 
  (\ref{dotQ}) and as a functional of   $Y=y-x$ and white noises.
  Then, making use that $Y(t)$ is an   Ornstein--Uhlenbeck process
  ($\dot{Y}=-2kY-\tau+\xi_{2}(t)-\xi_{1}(t)$),   $\langle Q \rangle$ and
  $\sigma^{2}$ can be evaluated from the moments and   correlations of $Y(t)$
  and $\xi(t)$.}   
\begin{equation}
\sigma^{2} = t_{0} \frac{k}{4} (T_{2}+ T_{1})^{2}.
\end{equation}
Note that the mean value and variance are extensive in $t_{0}$ and they are
also linear in $k$. The above results  yield to 
\begin{equation} \label{cxft}
Y(Q) \doteq \ln \frac{p(+Q)}{p(-Q)}=\Delta \beta Q (1-\gamma),
\end{equation}
where 
\begin{equation} 
\gamma= \left( \frac{T_{2}-T_{1}}{T_{2}+T_{1}} \right) ^{2}.
\end{equation}
This is one of the main results of this work. The expression does not depend
on any detail of the coupling mechanism.
\begin{figure}
\begin{center}
  \includegraphics[ angle=270, width=0.38\textwidth]{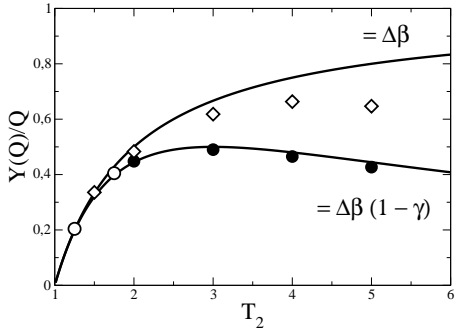}
  \caption{Prediction of the XFT (equation (\ref{xft!})) and the correction
    found due to the coupling 
    mechanism (equation (\ref{cxft})) in the Gaussian approximation. The
    symbols are obtained from numerical 
    data through different procedures (see text). The device considered is the
    passive Brownian transducer at $\tau=0$ and  $T_{1}=1$.}
   \label{corr}
\end{center}
\end{figure}
We obtain that the XFT holds for small $\Delta \beta$ ($T_{2} \simeq T_{1}$)
but an important correction of order $\Delta \beta ^{3}$ appears.
In Fig. \ref{corr} we plot the XFT (\ref{xft!}), the new prediction for the 
passive Brownian transducer (\ref{cxft}) and results from numerical
simulations, exploring the $T_2$ dependence.
The symbols correspond to numerical data and deserve a careful explanation. 
The circles are obtained at $kt_{0}=100$. For white circles, $Y(Q)$ has been
obtained by direct analysis of the probability distribution functions of
heat. Around  
$T_{2}>2$, negative values ($Q<0$) are rarer events and therefore
the tails are very difficult to observe directly. Black circles are
obtained assuming Gaussian behavior of the 
tails, though it is important to note that the distributions are
still checked to be fitted reasonably well by a Gaussian. The higher the
temperature $T_2$, the worse the Gaussian supposition is and so, for
$T_{2} \gtrsim 6$, we cannot conclude anything about $Y(Q)$.
Finally, diamonds correspond to data at $kt_{0}=10$ from which $Y(Q)$ is
directly measured from the histograms.
This case is also very  instructive because
the distributions observed are non-Gaussian (see Fig. \ref{one}).
The prediction (\ref{cxft}) has also been checked satisfactorily (data not
shown) in the nonlinear Brownian motor for $kt_{0}=100$,
$V_{0}=1$ and $d=12$. Then our simulations indicate that, despite the rapidly
increasing difficulty to test the theorem, important and systematic deviations
do appear. 

The $\gamma$ factor modifying the XFT is a signature of the
important role of the coupling between both systems.
This points out the applicability of the theorem for large $\Delta \beta$,
which strongly depends  
on the approximation made in the XFT derivation. It
consists on neglecting the interaction term coupling the two
bodies \cite{jar} by assuming that the energy involved in the coupling
mechanism is much smaller than the typical energy change in both systems. 
When comparing in our
model the typical energy of the coupling mechanism (the mean value of the
potential energy of the spring $\langle V_c \rangle =(T_2 + T_1)/4$), and the
typical energy change in every subsystem (the mean heat released 
$\langle Q \rangle$), for the parameter values $k=100$, $t_{0}=1$,
$T_{2}=5$ and $T_{1}=1$, we find $\langle V_c
\rangle / \langle Q \rangle \sim 0.0075$. Therefore, we must remark here 
that, albeit the interaction energy can be negligible, its consequences are
not.

\begin{figure}
\begin{center}
   \includegraphics[angle=270, width=0.38\textwidth]{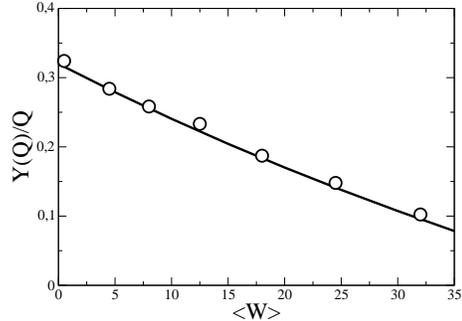}
   \caption{Numerical simulations (dots) and analytical prediction (line) of
     equation (\ref{tauxft}) for
     the relation of heat fluctuations versus the mean value of the work in the
     Brownian motor model defined in equations (\ref{123})--(\ref{124}). The
     agreement is very good, even for this nonlinear system. The data are
     obtained by direct measurement of the slope of $Y(Q)$.   
     The parameters used are $T_{1}=1$, $T_{2}=1.5$,
     $d=12$, $V_{0}=1$, $t_{0}=1$ and $k=100$.} 
  \label{Qf}
\end{center}
\end{figure}
The second point we study is the effect of an
external work into the system. We want to explore the possibility of an 
extension of the  relation (\ref{cxft}) for this case.
Taking  $\tau \neq 0 $ in  (\ref{123}), discarding the nonlinear ratchet
potential   
and proceeding as before for these type of calculations,
the quantity $\langle Q \rangle$ is    
\begin{equation}  \label{ac}
\langle Q \rangle =t_{0}\frac{1}{2}k(T_{2}-T_{1})- t_{0}\tau^{2}/4 =
\langle Q \rangle_c -\frac{\langle W \rangle}{2},
\label{QW}
\end{equation}
where $\langle Q \rangle_c$ is the mean heat conducted and
$\langle W \rangle= t_{0}\tau^{2}/2$ is the mean work. 
Remember that $\langle Q \rangle$ is the mean heat released by the heat bath
a $T_2$, so we are studying the fluctuations of this quantity, while heat 
exchanged at bath $T_1$ is different.  The
Fourier heat is conducted to the cold bath but also the hotter bath receives 
heat from the colder due to the external work. In
fact, each bath dissipates half of the total work. It is  
worth remarking that the sign of $\langle Q \rangle$ can be reversed for 
$\langle W \rangle > 2 \langle Q \rangle_c$, reversing the heat flux, now
from  
the cold bath to the hot one.
The calculus of the variance is more involved but we find
\begin{equation}  
\sigma^{2} = t_{0} \frac{k}{4}(T_{2}+T_{1})^{2}+ \langle W \rangle T_{1}.
\end{equation}
Using these results and defining the ratio of the two (model dependent)
energies involved, ${\cal R}=\langle W \rangle / \langle Q \rangle_{c}$,
the generalized relation for heat fluctuations is
\begin{equation} \label{tauxft}
\ln \frac{p(+Q)}{p(-Q)}=\Delta \beta Q (1-\gamma)\frac{1-{\cal R}
  /2}{1+f(T_1,T_2) {\cal R}}  
\end{equation}
where
\begin{equation}  
f(T_1,T_2)=\frac{2T_1(T_2-T_1)}{(T_2+T_1)^{2}}.
\end{equation}
The above extension depends on the 
mechanisms involved. Nevertheless, although, expression (\ref{tauxft}) has
been derived for a linear model, we can apply it to the nonlinear case
obtaining a very good agreement, as it is shown in Fig. \ref{Qf}. 
This means that we have found the terms that gather the most
relevant features and which work even for general nonlinear devices.
Notice that the torque used is in general beyond the stall
torque of the motor performance. This is so because for very small loads, thus
the ones that this motor is able to lift, it is very difficult to observe
changes in the distributions of $Q$. We must also stress that our analytical 
prediction does not involve any adjustable parameter and, as a consequence, it
could be confronted against other type of conducting and working devices.  

One can derive similar results if the heat 
interchanged by the cold bath at $T_1$ is considered instead of 
the heat transferred from the bath at $T_2$.
In this case, we have found (not shown 
here) that the relation (\ref{cxft}) is unchanged when 
we deal with the fluctuations of the energy dissipated in the cold bath.
Nevertheless for  the $\tau \neq 0$ case, the expressions vary but one can 
find the corresponding relations following the same type of
calculations. 
With respect to the conditions of our approach, we stress that it could be
possible to obtain analytical expressions for the statistical moments of $Q$ 
in the transient regime, after putting both systems initially in contact.
Transient corrections of order  $e^{-kt_{0}}$ appear  
which, in the limit $kt_{0}>>1$, can be discarded. Thus  the transient
regime is negligible compared to the steady state contribution. Therefore, all
the calculations in this letter are done in the steady 
state and in the long $t_0$ limit  (or big coupling).
This is, indeed, of great advantage because it makes it  possible to derive 
analytically the most dominant  contributions of the first
and second moments.
What is more, in such limit, although  $p(Q)$ is not rigorously
Gaussian because $Q(t)$ as defined in (\ref{dotQ}) is a nonlinear functional
of a Gaussian Orstein--Uhlenbeck process, our main results are dominated by
the Gaussian--like property of the distribution.
As a byproduct we have shown in a linear model that heat fluctuations
relative to the mean value $\sigma^2 / \langle Q \rangle$ are system
independent. It would be worthy to explore this result for other nonlinear
models. 

This work is the first application and test of the XFT \cite{jar} to
a non-ideal system, in the sense that the effects of the system-environment
coupling energy cannot  be neglected.
We have checked the sensitivity of the hypothesis of 
small interaction term in Ref. \cite{jar} for simple Brownian devices. Quite
surprisingly the XFT works 
better when the heat conducted is of the order of the 
energy stored in the coupling device but not when it is larger.
It is worth emphasizing that any attempt to write a model that
consists on two bodies interchanging heat needs a connection and,
therefore, the observations mentioned above are encountered. 
Hence, one can  neglect the energy stored in the connecting mechanism but its
effect has relevant consequences. 
However, in order to understand in more
detail the role of heat, work and coupling energy, it would be very
interesting to 
address these questions from a more general theoretical point of view.
The applicability of such results in theoretical models
and in experiments is of great importance for discovering and
understanding nonequilibrium statistical mechanics principles. 

Fruitful correspondence with Prof. Chris Jarzynski and valuable comments from
Prof. Chris Van den Broeck are gratefully acknowledged.
This research was supported by Ministerio de Educaci\'on y Ciencia (Spain)
under the  project BFM2003-07850 and the  grant  FPU-AP-2004-0770 (A. G--M.).

\end{document}